\documentclass[pra,aps,superscriptaddress,floatfix,tightenlines,twocolumn,showpacs]{revtex4}
\usepackage{amsmath}
\usepackage{amssymb}
\usepackage{amsfonts}
\usepackage{dcolumn}
\usepackage{epsfig}
\usepackage{subfigure}
\usepackage[dvips]{color}
\usepackage{bm}
\usepackage{times}
\usepackage{amsthm}
\usepackage{CJK}

\begin{document}
\title{Majorization relation in quantum critical systems}
\author{HUAI Lin-Ping}
\affiliation{Institute of Theoretical \& Computational Physics, School of Physics and Information Technology,
Shaanxi Normal University, Xi'an 710062, China}
\affiliation{Beijing National Laboratory for Condensed Matter Physics, Institute of Physics, Chinese Academy of Sciences, Beijing 100190, China}
\author{ZHANG Yu-Ran} 
\email{yrzhang@iphy.ac.cn}
\affiliation{Beijing National Laboratory for Condensed Matter Physics, Institute of Physics, Chinese Academy of Sciences, Beijing 100190, China}
\author{LIU Si-Yuan}
\affiliation{Institute of Modern Physics, Northwest University, Xi'an 710069, China}
\affiliation{Beijing National Laboratory for Condensed Matter Physics, Institute of Physics, Chinese Academy of Sciences, Beijing 100190, China}
\author{YANG Wen-Li}
\affiliation{Institute of Modern Physics, Northwest University, Xi'an 710069, China}
\author{QU Shi-Xian}
\email{sxqu@snnu.edu.cn}
\affiliation{Institute of Theoretical \& Computational Physics, School of Physics and Information Technology,
Shaanxi Normal University, Xi'an 710062, China}
\author{FAN Heng}
\email{hfan@iphy.ac.cn}
\affiliation{Beijing National Laboratory for Condensed Matter Physics, Institute of Physics, Chinese Academy of Sciences, Beijing 100190, China}
\affiliation{Collaborative Innovation Center of Quantum Matter, Beijing , China}
\pacs{64.60.A-;03.67.Mn;05.70.Jk}
\date{\today}

\begin{abstract}
The most basic local conversion is local operations and classical communications (LOCC),
which is also the most natural restriction in quantum information processing. We
investigate the conversions between the ground states in quantum critical systems
via LOCC and propose an novel method to reveal the different convertibility via majorization
relation when a quantum phase transition occurs. The ground-state local
convertibility in the one-dimensional transverse field Ising model is studied. It is shown
that the LOCC convertibility changes nearly at the phase transition point.
The relation between the order of quantum phase transitions and the LOCC convertibility
is discussed. Our results are compared with the corresponding results using the R\'{e}nyi
entropy and the LOCC convertibility with assisted entanglement.
\end{abstract}
\maketitle

Many developments in quantum-information processing (QIP) \cite{1} unveiling
the rich structure of quantum states and the nature of entanglement have offered many
insights into quantum many-body physics \cite{2}. Concepts from QIP have generated many
alternative indictors of quantum phase transitions, which has become a focus of attention in
detecting a number of critical points. For example, entanglements measured by concurrence
\cite{5}, negativity \cite{6}, geometric entanglement \cite{7}, and von Neumann entropy \cite{3,4}
have been investigated in several critical systems. From another viewpoint of quantum
correlations, other concepts in quantum information including mutual information \cite{9},
quantum discord \cite{8} and global quantum discord \cite{gqd}, have also been used for
detecting quantum phase transitions. Other investigations in entanglement spectra \cite{10,11,12}
and fidelity \cite{13} as well as the fidelity susceptibility of the ground state show their abilities
in exploring numerous phase transition points in various critical systems, as well. These
observations have achieved great success in understanding the deep nature of the different
phase transitions with the tools of quantum information science harnessed in the analysis of
quantum many-body systems. The reverse, however, seems unclear and tough.

\begin{figure}[b]
 \centering
\subfigure[]{\includegraphics[width=0.225\textwidth]{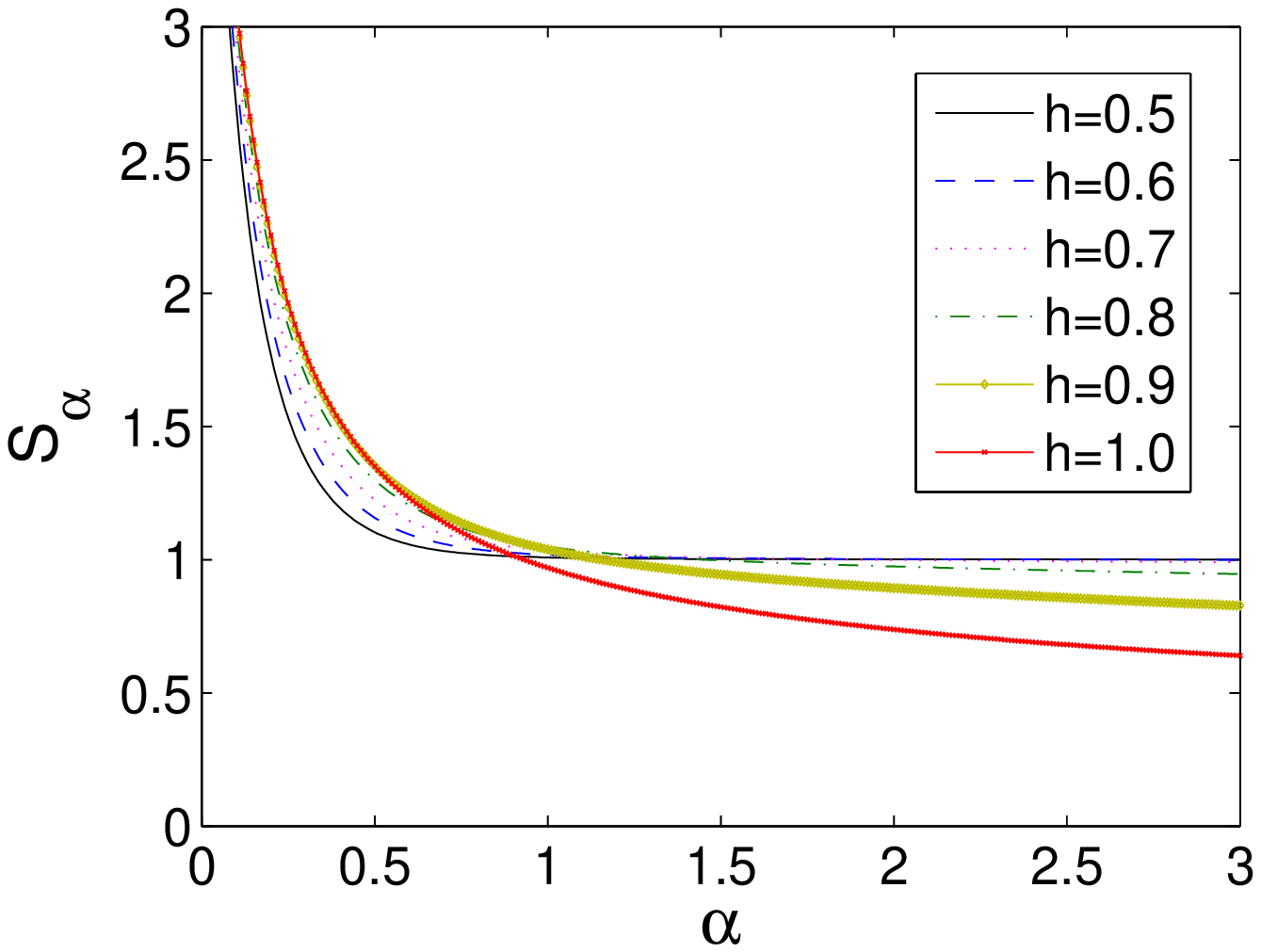}}
\subfigure[]{\includegraphics[width=0.225\textwidth]{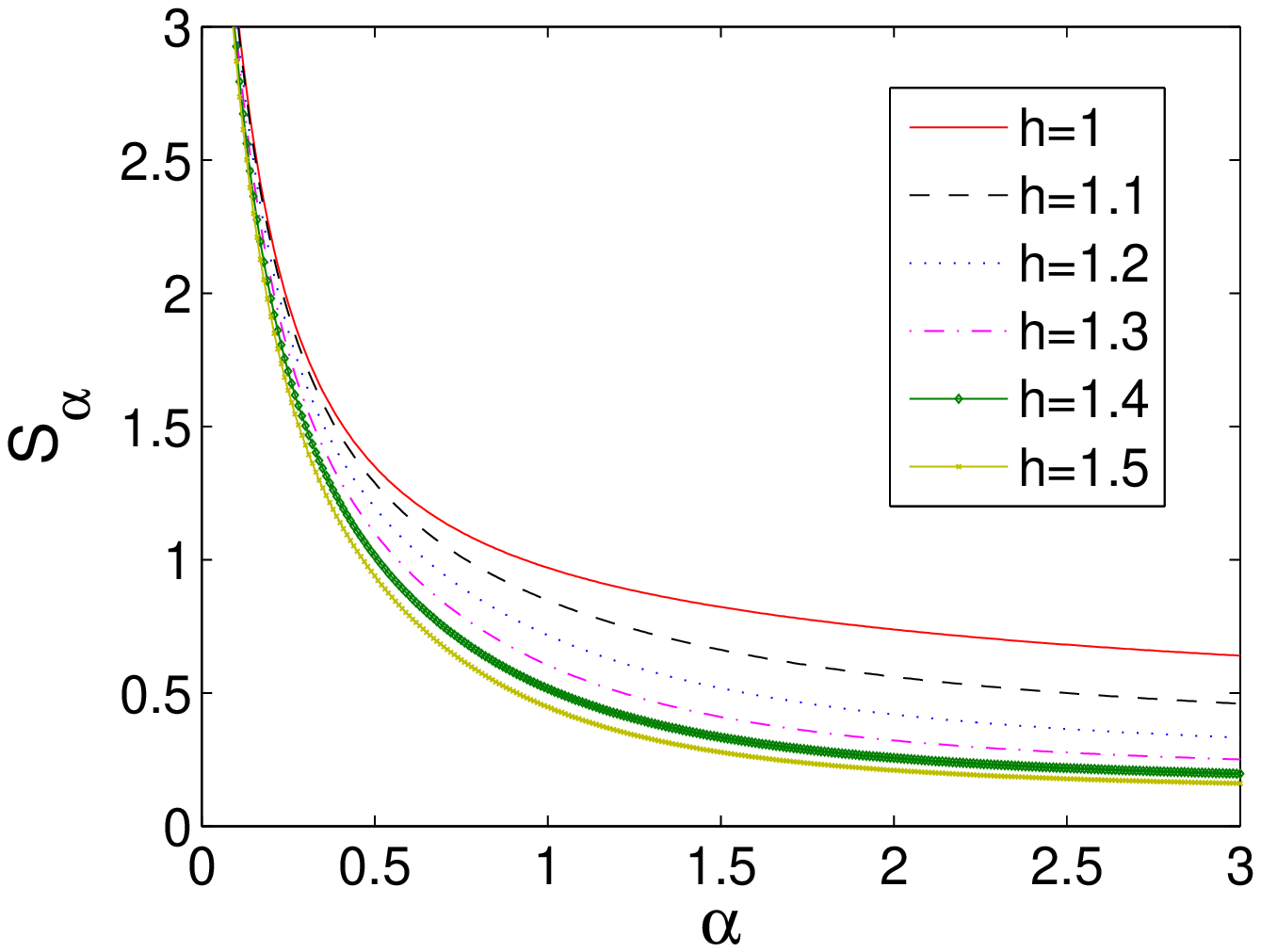}}
\caption{(Color online) R\'{e}nyi entropy for the ground state of the transverse field
Ising model v.s. $\alpha$. The R\'{e}nyi entropies are intercepted as (a) $h\leq1$, while they
are nonintercepted as (b) $h\geq1$, which gives a physical significance to QIP from quantum.
transitions}\label{f1}
\end{figure}

From a new point of view, Jian Cui and his collaborators reveal that the systems undergoing
quantum phase transition will also exhibit different operational properties from the perspective
of QIP \cite{cuipra,cuinc}. They demonstrate with several models that nearly at the critical points,
the entanglement-assisted local operations and classical communications (ELOCC) convertibility decided
via the R\'{e}nyi entropy interception changes suddenly, see Fig.~\ref{f1}. The R\'{e}nyi entropy
is defined as \cite{renyi}
\begin{eqnarray}
S_{\alpha}(\rho_{A})=\frac{1}{1-\alpha}\log_{2}\textrm{Tr}\rho_{A}^{\alpha}
\end{eqnarray}
and it tends to the von Neumann entropy as $\alpha\rightarrow1$. The unexpected results
suggest that not only are the tools of quantum information useful as alternative signatures of
quantum phase transitions, but  the study of quantum phase transitions may also offer
additional insight into QIP. However, the most basic and natural
local conversion is local operations and classical communications (LOCC), which has fabulous
applications in QIP, such as teleportation \cite{tele}, quantum states discrimination \cite{discr},
and quantum channel corrections \cite{correction}.

In this paper, we investigate the conversions
between the ground states in quantum critical systems via LOCC, and propose an novel method
to reveal the different convertibility via majorization relation \cite{maj} when a quantum phase
transition occurs. We study the one-dimensional transverse field Ising model and show that the
LOCC convertibility changes nearly at the phase transition point. Presenting the relation between
the order of quantum phase transitions and the LOCC convertibility, we compare our results with
the corresponding results using the ELOCC convertibility.

The majorization relations provide a necessary and sufficient condition
for the LOCC convertibility. For two bipartite pure states
$|\psi\rangle=\sum_{k=1}^{d}\sqrt{\lambda^{k}_{\psi}}|kk\rangle_{AB}$ and
$|\varphi\rangle=\ensuremath{\sum_{k=1}^{d}\sqrt{\lambda^{k}_{\varphi}}|kk\rangle}$,
if and only if the majorization relations
\begin{eqnarray}
\sum_{k=1}^{l}\lambda_{\psi}^{k}\geq\sum_{k=1}^{l}\lambda_{\varphi}^{k}
\end{eqnarray}
are satisfied for all $1\leq l\leq d$ (expressed as $\lambda_{\psi}\succ\lambda_{\varphi}$),
state $|\varphi\rangle$ can be transformed with 100\% probability of success to $|\psi\rangle$
by LOCC. Otherwise these two states are incomparable, i.e., $|\psi\rangle$ cannot be
converted to $|\varphi\rangle$ by LOCC, and vice versa. For example,
$|\psi_{1}\rangle=\sqrt{0.5}|00\rangle+\sqrt{0.3}|11\rangle+\sqrt{0.2}|22\rangle$,
$|\psi_{2}\rangle=\sqrt{0.5}|00\rangle+\sqrt{0.4}|11\rangle+\sqrt{0.1}|22\rangle$
and $|\psi_{3}\rangle=\sqrt{0.4}|00\rangle+\sqrt{0.4}|11\rangle+\sqrt{0.2}|22\rangle$.
It can be easily checked that $\lambda_{\psi_{1}}\prec\lambda_{\psi_{2}}$ which leads to
that $|\psi_{1}\rangle$ can be converted to $|\psi_{2}\rangle$ with certainty. For states
$|\psi_{1}\rangle$ and $|\psi_{3}\rangle$, no majorization relations can be fulfilled
(i.e. $\lambda_{\psi_{1}}\nprec\lambda_{\psi_{2}}$ and $\lambda_{\psi_{1}}\nsucc\lambda_{\psi_{2}}$)
and it indicates that neither state can be converted to the other with certainty.

\begin{figure}[t]
 \centering
\subfigure[]{\includegraphics[width=0.225\textwidth]{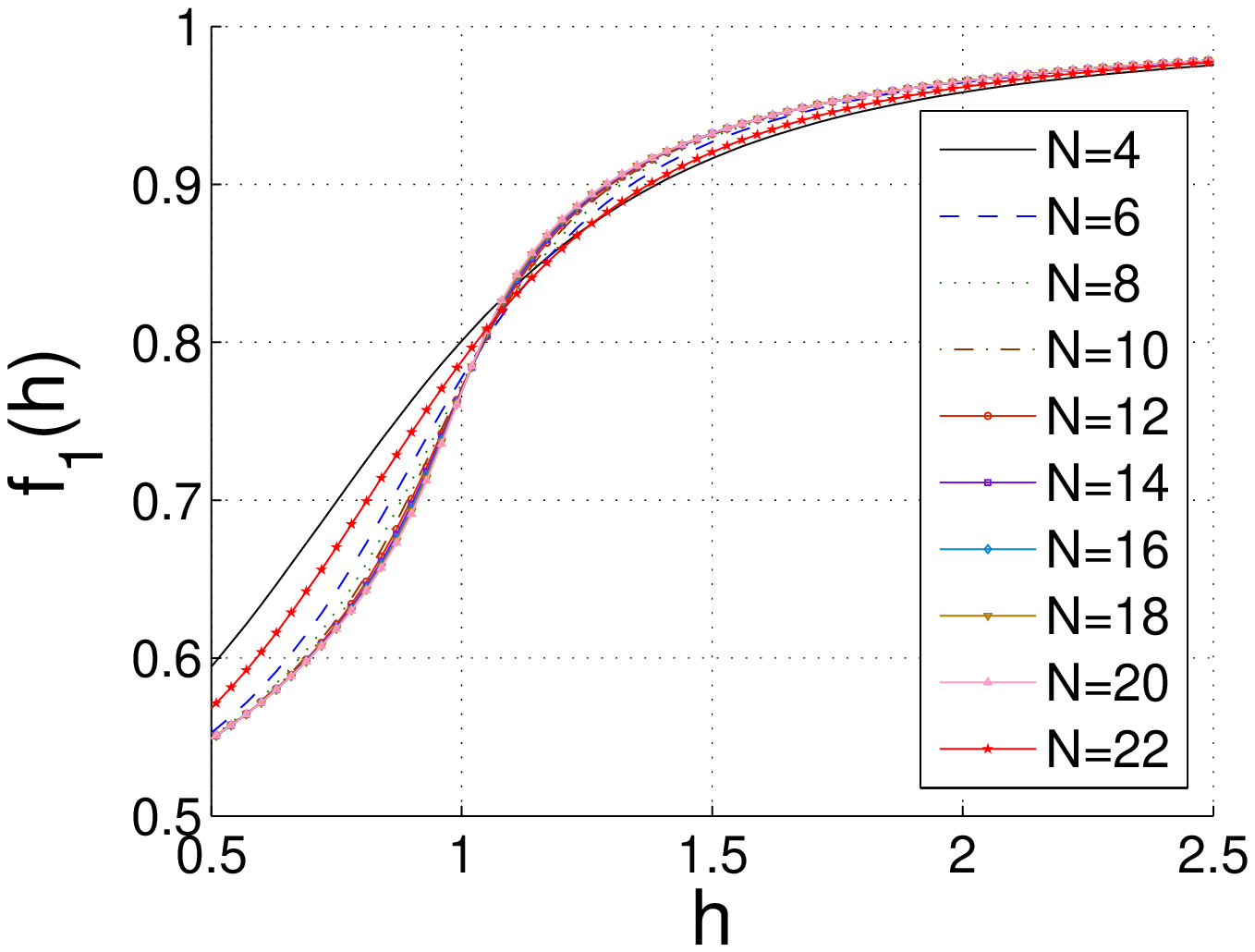}\label{f2a}}
\subfigure[]{\includegraphics[width=0.225\textwidth]{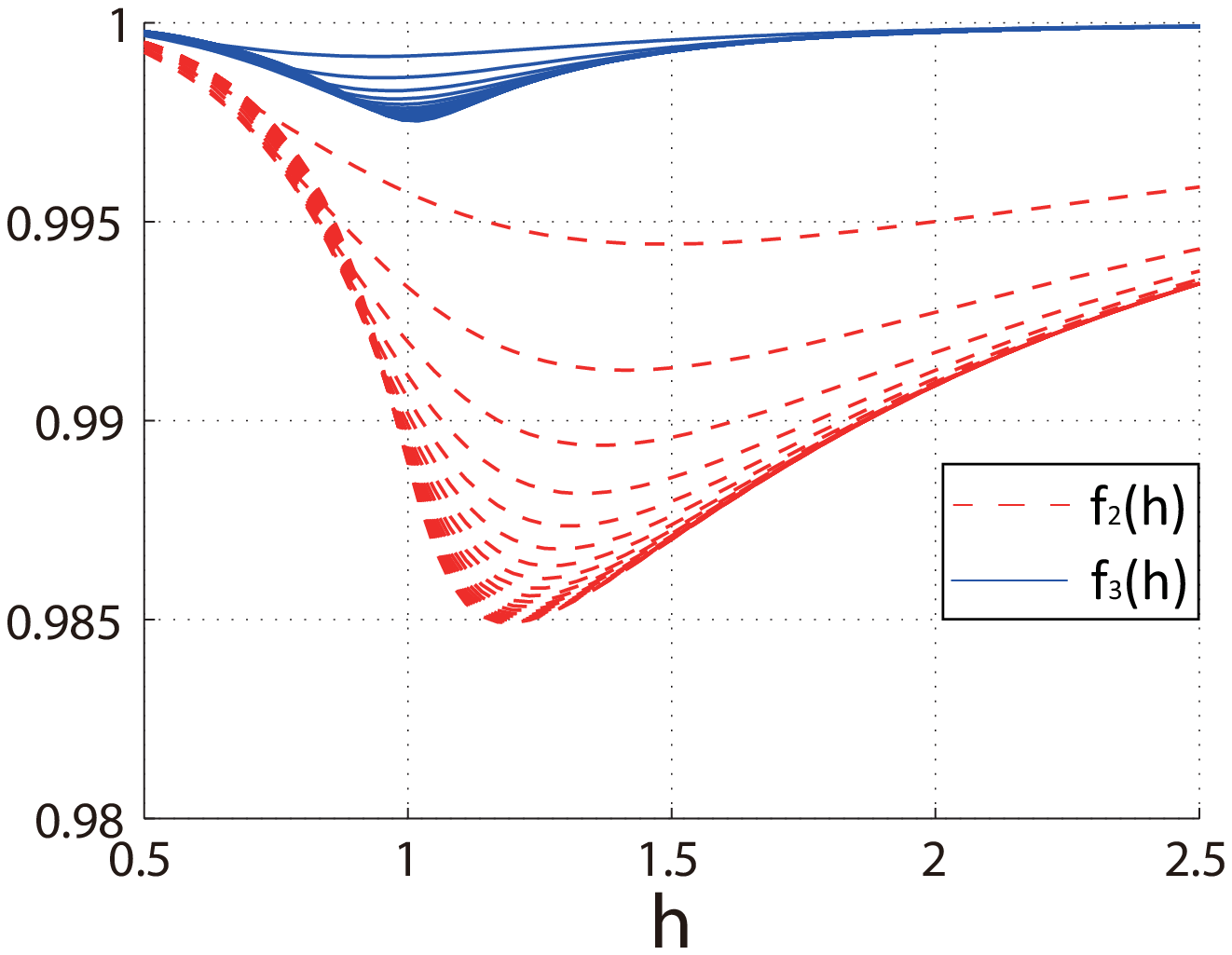}\label{f2b}}\\
\subfigure[]{\includegraphics[width=0.45\textwidth]{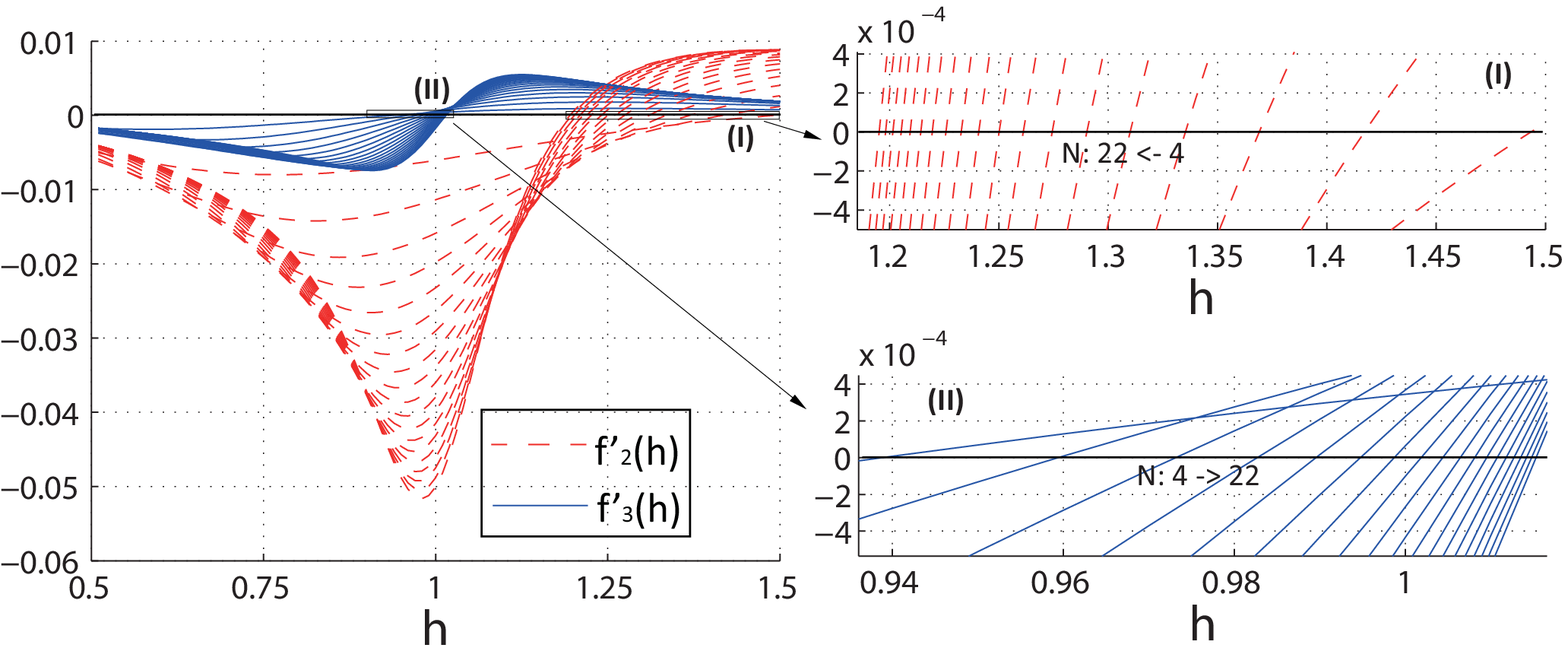}\label{f2c}}
\caption{(Color online) (a) Functions $f_{1}(h)$ against field parameter $h$ for Ising rings containing from $4$ to $22$
spins at zero temperature. For clearness, we merely perform the cases for even spin numbers. It is clear that $f_{1}(h)$
is monotonically increasing for all system sizes.
(b) We show $f_{2}(h)$ (dashed red lines) and $f_{3}(h)$ (solid blue lines) for different system sizes ranging from $4$ to $22$.
Both $f_{2}(h)$ and $f_{3}(h)$ firstly decrease and then increase.
(c) Derivatives $f'_{2}(h)$ (dashed red lines) and $f'_{3}(h)$ (solid blue lines)
against $h$ for $N$ ranging from $4$ to $22$.
The minimum points for $f_{2}(h)$ and $f_{3}(h)$ appear where $f'_{2}(h)=0$ and $f'_{3}(h)=0$.
}\label{f2}
\end{figure}

Next, given the transverse field Ising model in the zero-temperature case, we use majorization
relations to investigate the relationship between the critical points and the LOCC convertibility
of the ground state. The Hamiltonian for a chain of $N$ spin-$\frac{1}{2}$ particles reads
\begin{eqnarray}
H=-\sum_{i=1}^{N}\left(\sigma_{i}^{x}\sigma_{i+1}^{x}+h\sigma_{i}^{z}\right),
\label{ham}
\end{eqnarray}
where periodic boundary conditions are assumed: $N+1\rightarrow1$. $\sigma_{i}^{x}$ and
$\sigma_{i}^{z}$ stand for the Pauli matrices, $h$ is the field parameter. A second-order quantum
phase transition takes place at $h=1$. For $h>1$, it is the paramagnetic phase, while it is
the ferromagnetic phase for $0<h<1$. We study the ground states of this model for system sizes
$N=4,\cdots,22$ with the field parameter $h$ varying from $0.5$ to $2.5$. The ground states
labelled as $|G(h)\rangle_{AB}$ are obtained by exactly diagonalizing the whole Hamiltonian (\ref{ham}).
This new proposal is also worth further investigation by other numerical methods such as Lanczos algorithm and density matrix renormalization group (DMRG), which are not included in this paper.

We consider three largest eigenvalues of the reduced states of any two neighbor spins as $\lambda_{1}(h)$,
$\lambda_{2}(h)$, and $\lambda_{3}(h)$ in descending order. In order to
detect the majorization relations between two ground states $|G(h)\rangle_{AB}$ and
$|G(h+\Delta)\rangle_{AB}$ given some infinitesimal $\Delta$, we should judge the
monotonicities of three functions for the field parameter $h$:
\begin{eqnarray}
f_{1}(h)&\equiv&\lambda_{1}(h),\\
f_{2}(h)&\equiv&\lambda_{1}(h)+\lambda_{2}(h),\\
f_{3}(h)&\equiv&\lambda_{1}(h)+\lambda_{2}(h)+\lambda_{3}(h).
\end{eqnarray}
Therefore, three cases will be met: $(i)$ when monotonicities of these three functions are all
non-increasing,  $|G(h+\Delta)\rangle_{AB}$ can be converted to $|G(h)\rangle_{AB}$ by LOCC with
certainty; $(ii)$ when monotonicities of these three functions are all non-decreasing
$|G(h)\rangle_{AB}$ can be converted to $|G(h+\Delta)\rangle_{AB}$ by LOCC with certainty; $(iii)$
except for these two cases,  neither state can be converted to the other via LOCC with certainty.

\begin{figure}[t]
 \centering
\includegraphics[width=0.45\textwidth]{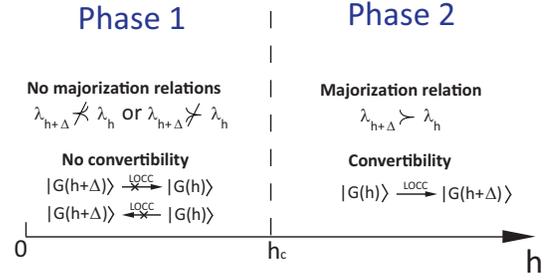}\label{f2d}
\caption{(Color online) Different LOCC convertibilities for two phases  in the transverse field Ising model.
}\label{fa}
\end{figure}

Function $f_{1}(h)$ against the field parameter for system size $N$ varying from $4$ to $22$
is shown in Fig.~\ref{f2a}, where we find $f_{1}(h)$ is monotonically increasing for $h\in[0.5,2.5]$.
In Fig.~\ref{f2b}, for system sizes ranging from $4$ to $16$,  dashed red lines and solid blue lines
denote to
$f_{2}(h)$ and $f_{3}(h)$ versus $h$, respectively. We also present the derivatives of $f_{2}(h)$ and
$f_{3}(h)$ in Fig.~\ref{f2c}. Both functions firstly decrease and then increase
starting nearly at the points $h^{c}\simeq1$, where $f'_{2}(h)f'_{3}(h)=0$, $f'_{2}(h)\geq0$ and $f'_{3}(h)\geq0$ are fulfilled.
Thus, we can conclude for the transverse field Ising model that
the majorization relations $\lambda_{h+\Delta}\succ\lambda_{h}$ hold when $h>h^{c}$,
and meanwhile, $|G(h)\rangle_{AB}$ can
be transformed to $|G(h+\Delta)\rangle_{AB}$ by LOCC with certainty. For the rest range of $h$, no
majorization relation can be found. These results indicate that there is a distinct change in the
properties of ground states at the critical points $h^{c}$, see Fig.~\ref{fa}. It seems credible to detect
the phase transition points via the majorization relations for the LOCC convertibility, and in contrary,
it suggests that the  LOCC convertibilities of the ground states of transverse field Ising model in two
phases are different.

\begin{figure}[t]
 \centering
\subfigure[]{\includegraphics[width=0.22\textwidth]{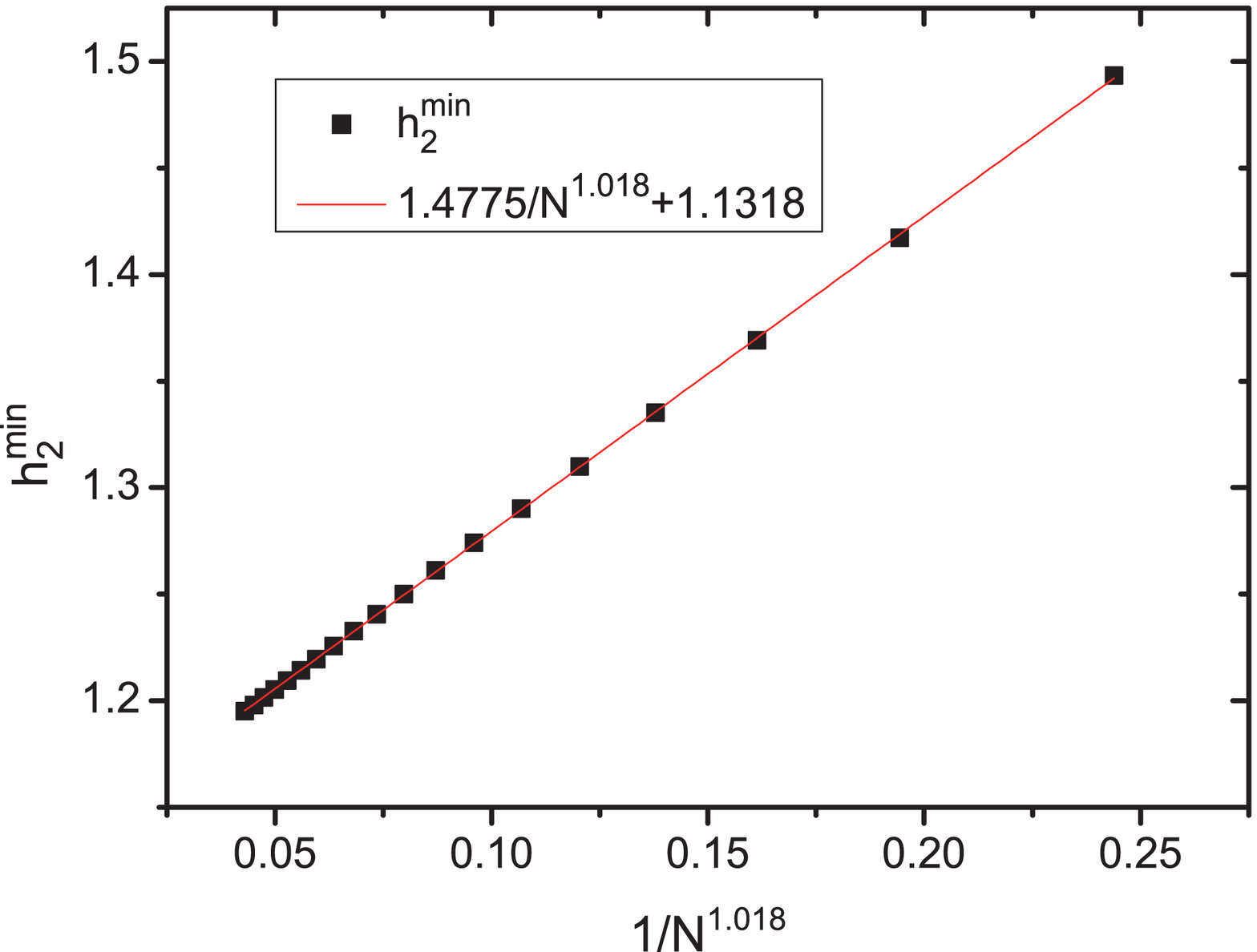}\label{f3a}}
\subfigure[]{\includegraphics[width=0.22\textwidth]{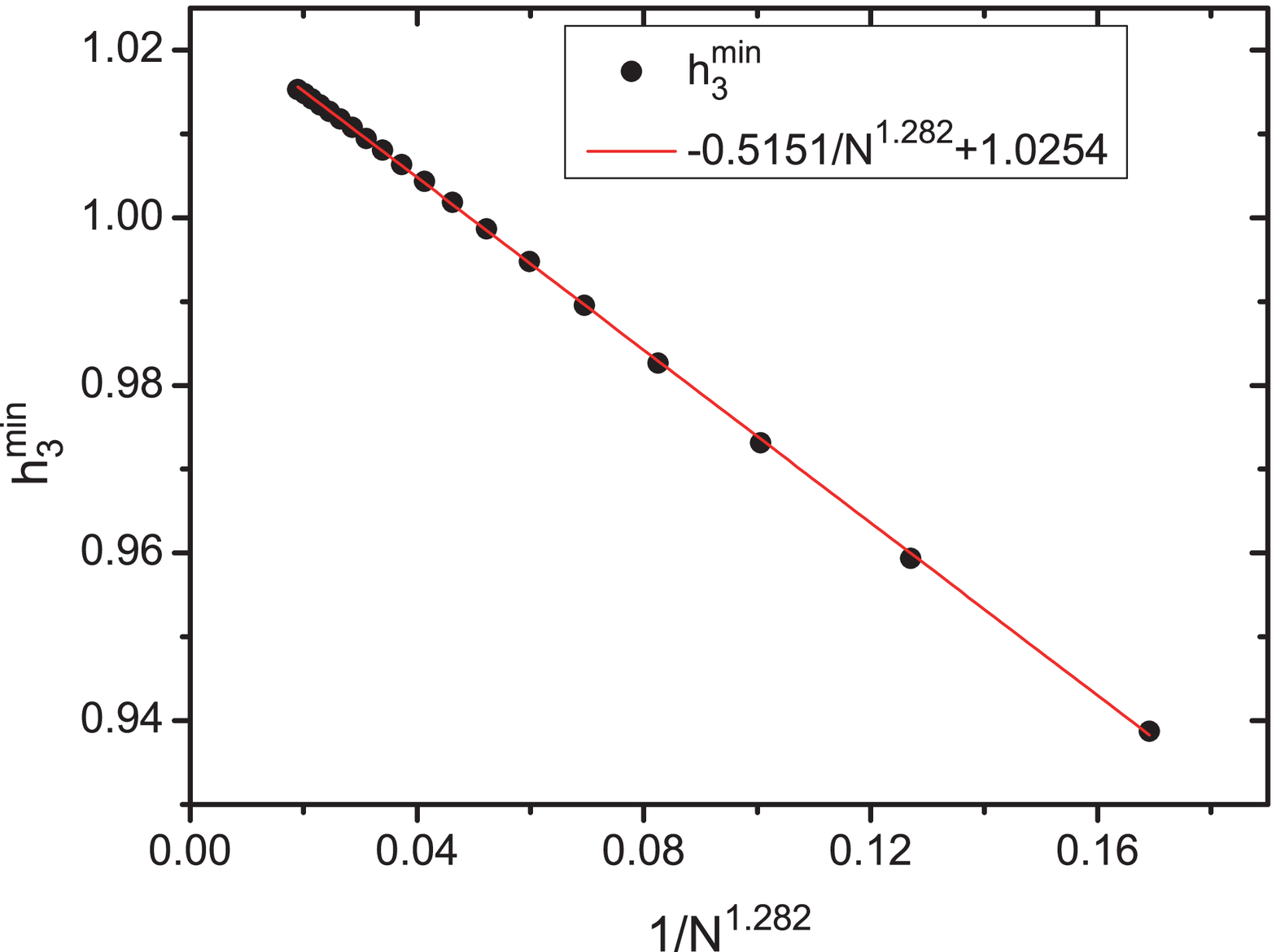}\label{f3b}}
\caption{Finite-size scaling  analysis of the minimum points of
$f_{2}(h)$ and $f_{3}(h)$ in Ising model: (a) is for $h^{min}_{2}$ and (b) is for $h_{3}^{min}$}\label{f3}.
\end{figure}

To get rid of the finite-size effect, we give a scaling analysis of the critical points $h^{c}$ in
Fig.~\ref{f3}. The parameters $h^{min}_{2}$ and $h^{min}_{3}$ for the minimum points of
$f_{2}(h)$ and $f_{3}(h)$ for different system sizes  are plotted in Fig.~\ref{f3a} and \ref{f3b},
respectively. They can be fitted as
$h^{min}_{2}=1.4775/N^{1.018}+1.1318$ with accuracy $3.24\times10^{-4}$ and
$h^{min}_{3}=-0.5151/N^{1.282}+1.0254$
with accuracy $1.07\times10^{-4}$. Therefore, the critical point for large $N$ should be chosen as
$h^{c}=1.1318$ on the basis of the majorization relations and the LOCC convertibility. As $N\rightarrow\infty$,
it can clearly demonstrate that when $h\geq1.1318$, $|G(h)\rangle_{AB}$ can be transformed
to $|G(h+\Delta)\rangle_{AB}$ by LOCC with certainty, and in the region $0<h<1.1318$, neither state
can be converted to the other via LOCC with certainty.

Then we compare our results with the results obtained by the ELOCC convertibility with
R\'{e}nyi entropy. States $|\psi\rangle_{AB}$ can be transformed to $|\psi'\rangle_{AB}$
via ELOCC if and only if their R\'{e}nyi entropies satisfy $S_{\alpha}(\rho_{A})\geq S_{\alpha}(\rho_{A}')$
for all $\alpha$ given $\rho_{A}$ and $\rho_{A}'$ the reduced density matrices of $|\psi\rangle_{AB}$
and $|\psi'\rangle_{AB}$, respectively \cite{ea}. In Ref.~\cite{cuipra}, it shows that the phase
transition point of the transverse field Ising model via the ELOCC proposal with R\'{e}nyi entropy is
$h^{c}_{E}=0.9940$ as $N\rightarrow\infty$ which offers a higher accuracy for detecting the critical
points for the transverse field Ising model. It demonstrates that the point where the ELOCC convertibility
suddenly changes stands closer to the critical points than the one at which the LOCC convertibility
changes. Combined with this conclusion, a clearer and further description of the local convertibility
of the ground states of the transverse field Ising model can be presented: In the region $0<h<0.9940$ within the
ferromagnetic phase, neither LOCC nor ELOCC convertibilities exist, in the region $h>1.1318$
within the paramagnetic phase, both LOCC and ELOCC convertibilities exist, and in the small interval
$0.994<h<1.1318$ around the critical point $h=1$, merely the ELOCC convertibility exists.

In conclusion, we investigate the majorization relations and the LOCC convertibility in quantum
critical systems. We develop a proposal to describe the LOCC conversion properties of quantum
critical systems by examining the majorization relations. We apply this proposal to  study
one-dimensional transverse field Ising model, which shows that the LOCC convertibility changes
at $h^{c}=1.1318$ nearly at the critical point. ELOCC convertibility and LOCC convertibility in most
areas of the paramagnetic phase are both stronger than those in the
ferromagnetic phase, however, in small intervals around the critical ponit $h=1$, merely the ELOCC
convertibility exists. In the ferromagnetic phase where $0<h<1$, the ground states can nearly convert
to each other by LOCC or ELOCC; whereas in the paramagnetic phase where $h>1$, the situation is
slightly more complicated. The LOCC convertibility via majorization relations applied for quantum phase transition is a new method. It will help us understand from a different view point the complicated phenomena of quantum critical systems. This LOCC proposal in detecting quantum phase transition can play a complementary role to the ELOCC method; it should be applicable in other systems similar as in the one-dimensional transverse field Ising model. This paper will
enlighten extensive studies of quantum phase transitions from the perspective of local convertibility,
e.g., finite-temperature phase transitions and other quantum many-body models.

\begin{acknowledgments}
We thank Jun-Peng Cao, Dong Wang and Yu Zeng for valuable discussions.
This work is supported by NSFC (11175248), ¡°973¡±
program (2010CB922904).
\end{acknowledgments}

\end{document}